\documentclass[letterpaper,aps,prl,twocolumn]{revtex4}
\usepackage{graphicx,bm}
\usepackage{amsmath,amssymb}

\begin{document}

\title{Anomalous isotope effect in phonon-dressed iron-based superconductors}
\author{Wen-Min Huang$^{1}$, Hong-Yan Shih$^{2}$, Fa Wang$^3$ and Hsiu-Hau Lin$^{4}$}
\affiliation{
$^1$Physics Division, National Center for Theoretical Sciences, Hsinchu 300, Taiwan\\$^2$Department of Physics, University of Illinois at Urbana-Champaign, Urbana, Illinois 61801-3080, USA\\
$^3$Department of Physics, Massachusetts Institute of Technology, Cambridge, MA 02139, USA\\
$^4$Department of Physics, National Tsing Hua University, Hsinchu 300, Taiwan}
\date{\today}

\begin{abstract}
The role of electron-phonon interactions in iron-based superconductor is currently under debate with conflicting experimental reports on the isotope effect. To address this important issue, we employ the renormalization-group method to investigate the competition between electron-electron and electron-phonon interactions in these materials. The renormalization-group analysis shows that the ground state is a phonon-dressed unconventional superconductor: the dominant electronic interactions account for pairing mechanism while electron-phonon interactions are subdominant. Because of the phonon dressing, the isotope effect of the critical temperature can be normal or reversed, depending on whether the retarded intra- or inter-band interactions are altered upon isotope substitutions. The connection between the anomalous isotope effect and the unconventional pairing symmetry is discussed at the end.
\end{abstract}

\maketitle

What role phonons play for electron pairing always stimulates intense discussions when a new superconductor is discovered\cite{Norman2011} such as cuprates\cite{Kresin}, magnesium diboride\cite{Canfield}, sodium cobalt oxide\cite{Donkov} and recently found iron-based superconductors\cite{Paglione,Stewart2011}.
The first checking point is the critical temperature of superconductivity.
By studying density of states for phonons, the critical temperature can be estimated from the Bardeen-Cooper-Schrieffer (BCS) theory or the Migdal-Eliashberg formula and then compared with the experimental data to see the deviation from conventional superconductivity\cite{Kortus, Rueff, Singh, Christianson, Boeri}.
In unconventional superconductors, the phonon-mediated interactions are insufficient to explain the pairing mechanism and it is of crucial importance to study the interplay between electron-electron and electron-phonon interactions\cite{Iliev,Lupi, Uchiyama,Jinho,Granath,Tacon}.
For instance, even when the pairing mechanism is electronic origin, dispersions observed in angle-resolved photoemission spectroscopy manifest distortions upon isotope substitutions\cite{Lanzara, Hasan, Yang, Gweon, Giustino,Iwasawa, Richard}.

To reveal the role of phonons, isotope effect in the superconducting temperature $T_c$ is commonly used\cite{Kresin, Canfield, Budko, Yokoi, Liu, Shirage,Shirage2010,Khasanov2010}. By detecting the change in $T_c$ upon isotope substitutions, the isotope exponent $\alpha = - d(\log T_c)/d(\log M)$ can be obtained\cite{Yanagisawa2009,Bussmann-Holder2011}, where $M$ is the mass of the substituted element. In conventional superconductors where electron-phonon interactions reign, the exponent is $\alpha =1/2$. On the other hand, in an unconventional superconductor without any relevant phonon-mediated interactions, $\alpha=0$ is expected. The isotope effect observed in iron-based superconductor\cite{Granath,Tacon,Richard,Liu,Shirage,Shirage2010,Khasanov2010} seems to tell a more complicated story. For instance, a strong isotope effect by iron substitution\cite{Liu} is found in SmFeAsO$_{1 - x}$F$_x$ and Ba$_{1 - x}$K$_x$Fe$_2$As$_2$, almost as large as that in conventional superconductors. On the contrary, inverse isotope effect\cite{Shirage} is also spotted in (Ba,K)Fe$_{2}$As$_{2}$ with different isotope substitutions.

Motivated by the controversy, we investigate the competition between electron-electron and electron-phonon interactions by the unbiased renormalization-group (RG) method. Due to the retarded nature of the phonon-mediated interactions, the energy dependence must be included. The minimal approach to include both simultaneous and retarded interactions can be accomplished by the step-shape approximation\cite{Zimanyi,Seidel} as shown in Fig. 1(a),   
\begin{eqnarray}\label{g}
g_i(\omega)=g_i +\tilde{g}_i \Theta(\omega_D-\omega),
\end{eqnarray}
where $g_i$ and $\tilde{g}_i$ represent (instantaneous) electronic interactions and (retarded) phonon-mediated ones. The energy scale for the retarded interactions is set by the Debye frequency $\omega_D$. Our RG analysis reveals that the pairing mechanism is dominated by the electronic interactions $g_i$. But, the retarded interactions $\tilde{g}_i$ also grow under RG transformation and become relevant in low-energy limit. Inclusion of these subdominant interactions leads to anomalous isotope effect. Though a quantitative estimate of the isotope exponent $\alpha$ is difficult in the presence of both types of interactions, we demonstrate how the exponent can be extracted numerically from RG flows in weak coupling. Surprisingly, the sign of the exponent $\alpha$ sensitively depends on whether the inter- and/or intra-band interactions are altered by isotope substitutions.

\begin{figure}
\centering
\includegraphics[width=8.3cm]{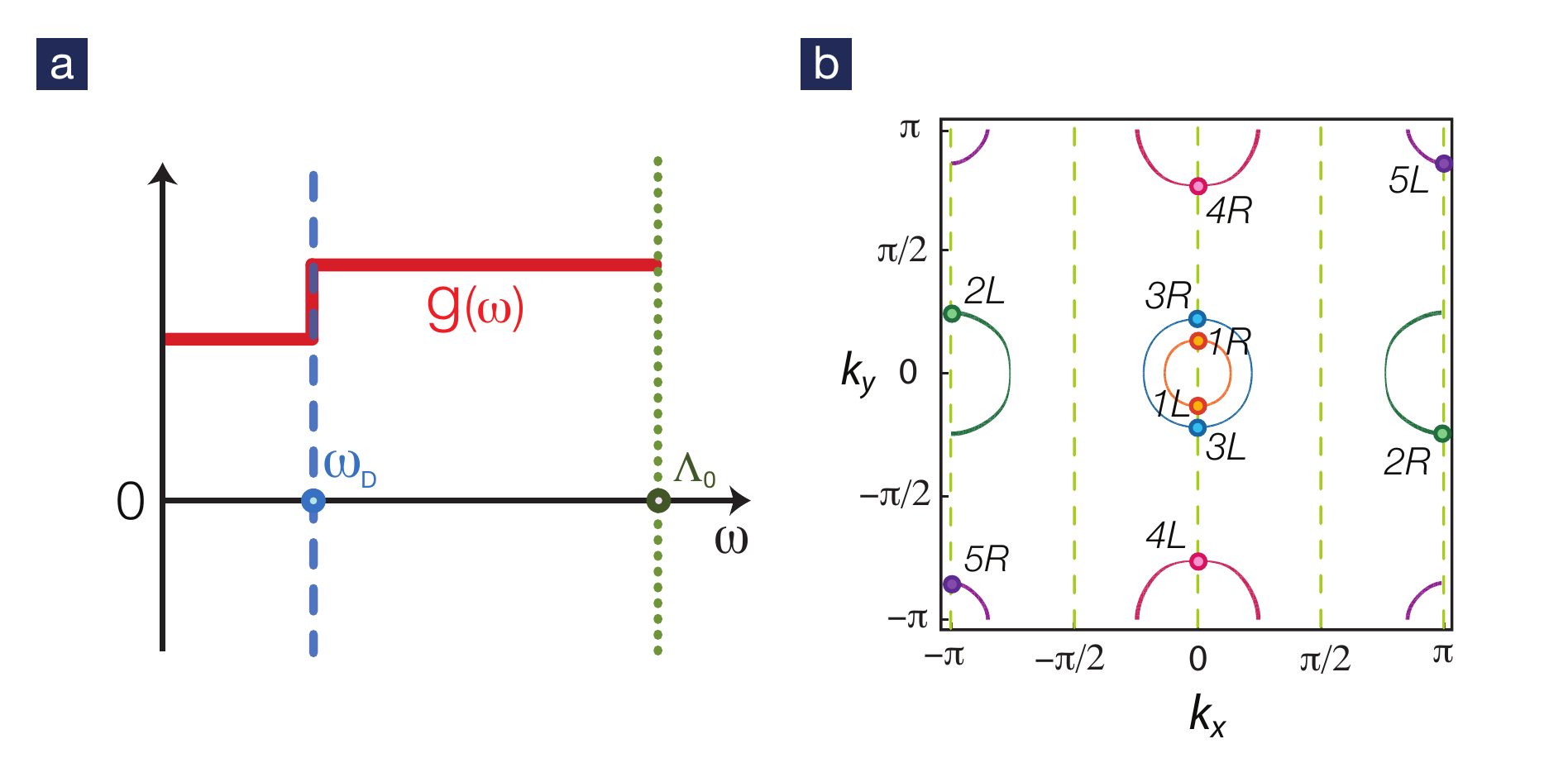}
\caption{(a) Step-like interaction profile for simultaneous and retarded interactions. A sharp step is assumed at the Debye frequency $\omega_D$. (b) Fermiology of the five-band model $x=0.1$. These Fermi surfaces are well sampled by five pairs of Fermi points, equivalent to the four-leg geometry with quantized momenta (dashed lines).
 }
\label{FS}
\end{figure}
\begin{figure}
\centering
\includegraphics[width=8.cm]{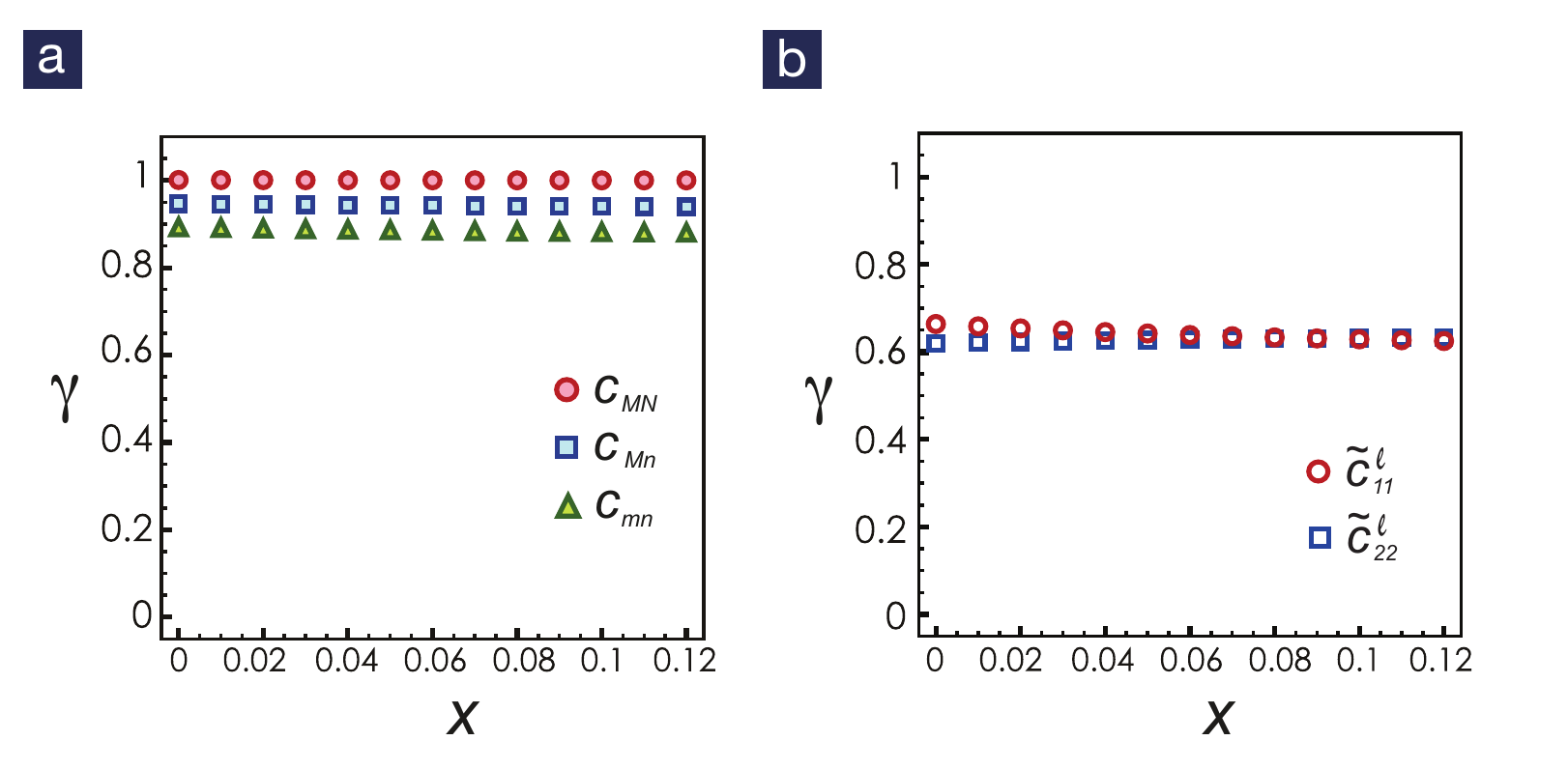}
\caption{RG exponents for (a) the simultaneous and (b) the retarded Cooper scatterings. The dominant interactions are pairing hopping between and within band 1 and band 2, with maximal exponent of unity, while other relevant couplings are subdominant with RG exponent smaller than one.
}
\end{figure}

To illustrate how the RG scheme works, we start with a five-orbital tight-binding model for iron-based superconductors with generalized on-site interactions,
\begin{eqnarray}\label{model}
H &=& \sum_{\bm{p},a,b}\sum_{\alpha}
c^{\dag}_{\bm{p}a \alpha}K_{ab}(\bm{p})c_{\bm{p}b\alpha}
+ U_1\sum_{i,a}n_{ia\uparrow}n_{ia\downarrow}
\nonumber\\
&+&U_2\sum_{i,a<b}\sum_{\alpha,\beta}n_{ia\alpha}n_{ib\beta}
+J_H\sum_{i,a<b}\sum_{\alpha,\beta}
c^{\dag}_{ia\alpha}c_{ib\alpha}c^{\dag}_{ib\beta}c_{ia\beta}
\nonumber\\
&+&J_H\sum_{i,a<b} \left[ c^{\dag}_{ia\uparrow}c^{\dag}_{ia\downarrow}
c_{ib\downarrow} c_{ib\uparrow}+{\rm H.c.}\right] \bigg\},
\end{eqnarray}
where $a,b=1,2,...,5$ label the five $d$-orbitals of Fe, $1:d_{3Z^2-R^2}$, $2:d_{XZ}$, $3:d_{YZ}$, $4:d_{X^2-Y^2}$, $5:d_{XY}$, and $\alpha=\uparrow, \downarrow$ is the spin index. The kinetic matrix $K_{ab}$ in the momentum space has been constructed in previous studies\cite{Kuroki}. The generalized on-site interactions consist of three parts: intra-orbital $U_1$, inter-orbital $U_2$ and Hund's coupling $J_H$. Adopted from previous studies, we choose the values, $U_1= 4$ eV, $U_2=2$ eV and $J_H=0.7$ eV for numerical studies here.

Fermiology is important in the multi-band superconductors. The electron doping $x$ is related to the band filling $n=6+x$ ($n=10$ for completely filled bands) here and the Fermi surface at$x=0.1$ is illustrated in Fig. 1(b). There are five active bands: two hole pockets centered at $(0,0)$ and another hole pocket centered at $(\pi,\pi)$ while two electron pockets located at $(\pi,0)$and $(0,\pi)$ points\cite{Ding}. To simplify the RG analysis, we sample each pocket with one pair of Fermi points (required by time-reversal symmetry). This is equivalent to a four-leg ladder geometry with quantized momenta as shown in Fig. 1(b). In the low-energy limit, the effective Hamiltonian\cite{Balents96,Lin97,Lin98a} is captured by five pairs of chiral fermions with different velocities. 

The interactions between these chiral fermions fall into two categories\cite{Lin05,Seidel}: Cooper scattering $c^l_{ij}, c^s_{ij}$ and forward scattering $f^l_{ij}, f^s_{ij}$. The retarded ones share the same classification, denoted with an extra tilde symbol. The RG equations for the simultaneous interactions are, 
\begin{eqnarray}\label{sRGeq}
\nonumber\dot{c}^l_{ii}&&\hspace{-0.3cm}=-2\sum_{k\neq i}\alpha_{ii,k}c^l_{ik}c^s_{ki}-2\left(c^l_{ii}\right)^2,\\
\nonumber\dot{c}^s_{ii}&&\hspace{-0.3cm}=-\sum_{k\neq i}\alpha_{ii,k}\left[\left(c^l_{ik}\right)^2+\left(c^s_{ik}\right)^2\right]-\left(c^l_{ii}\right)^2,\\
\nonumber\dot{c}^l_{ij}&&\hspace{-0.3cm}=-\sum_{k}\alpha_{ij,k}\left[c^l_{ik}c^s_{kj}+c^l_{jk}c^s_{ki}\right]-4f_{ij}^lc_{ij}^l\\
\nonumber&&\hspace{0.6cm}+2f_{ij}^lc_{ij}^s+2f_{ij}^sc_{ij}^l,\\
\nonumber\dot{c}^s_{ij}&&\hspace{-0.3cm}=-\sum_{k}\alpha_{ij,k}\left[c^l_{ik}c^l_{kj}+c^s_{ik}c^s_{kj}\right]+2f_{ij}^sc_{ij}^s,\\
\nonumber\dot{f}^l_{ij}&&\hspace{-0.3cm}=-2\left(f_{ij}^l\right)^2-2\left(c_{ij}^l\right)^2+2c_{ij}^lc_{ij}^s,\\
\dot{f}^s_{ij}&&\hspace{-0.3cm}=\left(c_{ij}^s\right)^2-\left(f_{ij}^l\right)^2,
\end{eqnarray}
where $\dot{g}=dg/dl$, where $l=\ln(\Lambda_0/\Lambda)$ is the logarithm of the ratio between bare energy cutoff $\Lambda_0$ and the running cutoff $\Lambda$. The tensor $\alpha_{ij,k}=(v_i+v_k)(v_j+v_k)/[2v_k(v_i+v_j)]$ with $v_i$ representing the Fermi velocities.

The second set of equations describes how the retarded interactions are renormalized,
\begin{eqnarray}\label{rRGeq}
\nonumber\dot{\tilde{c}}^l_{ii}&&\hspace{-0.3cm}=2\tilde{c}^l_{ii}c^s_{ii}-4\tilde{c}^l_{ii}c^l_{ii}-2\left(\tilde{c}^l_{ii}\right)^2,\\
\nonumber\dot{\tilde{c}}^l_{ij}&&\hspace{-0.3cm}=-4\tilde{f}^l_{ij}c^l_{ij}-4f^l_{ij}\tilde{c}^l_{ij}-4\tilde{f}^l_{ij}\tilde{c}^l_{ij}+2\tilde{f}^l_{ij}c^s_{ij}+2f^s_{ij}\tilde{c}^l_{ij},\\
\nonumber\dot{\tilde{f}}^l_{ij}&&\hspace{-0.3cm}=-4\tilde{f}^l_{ij}f^l_{ij}-2\left(\tilde{f}^l_{ij}\right)^2-4\tilde{c}^l_{ij}c^l_{ij}-2\left(\tilde{c}^l_{ij}\right)^2\\
&&\hspace{0.3cm}+2\tilde{f}^l_{ij}f^s_{ij}+\tilde{c}^l_{ij}c^s_{ij}.
\end{eqnarray}
Note that we separate the intra-band and inter-band couplings for clarity, i.e. $i \neq j$ in the above RG equations. In fact, the separation is necessary because we shall see later that inter-band and intra-band couplings play different roles in the low-energy limit. In addition, $f_{ii}=0$ and $\tilde{f}_{ii}=0$ to avoid double counting. 

\begin{figure}
\centering
\includegraphics[width=8.cm]{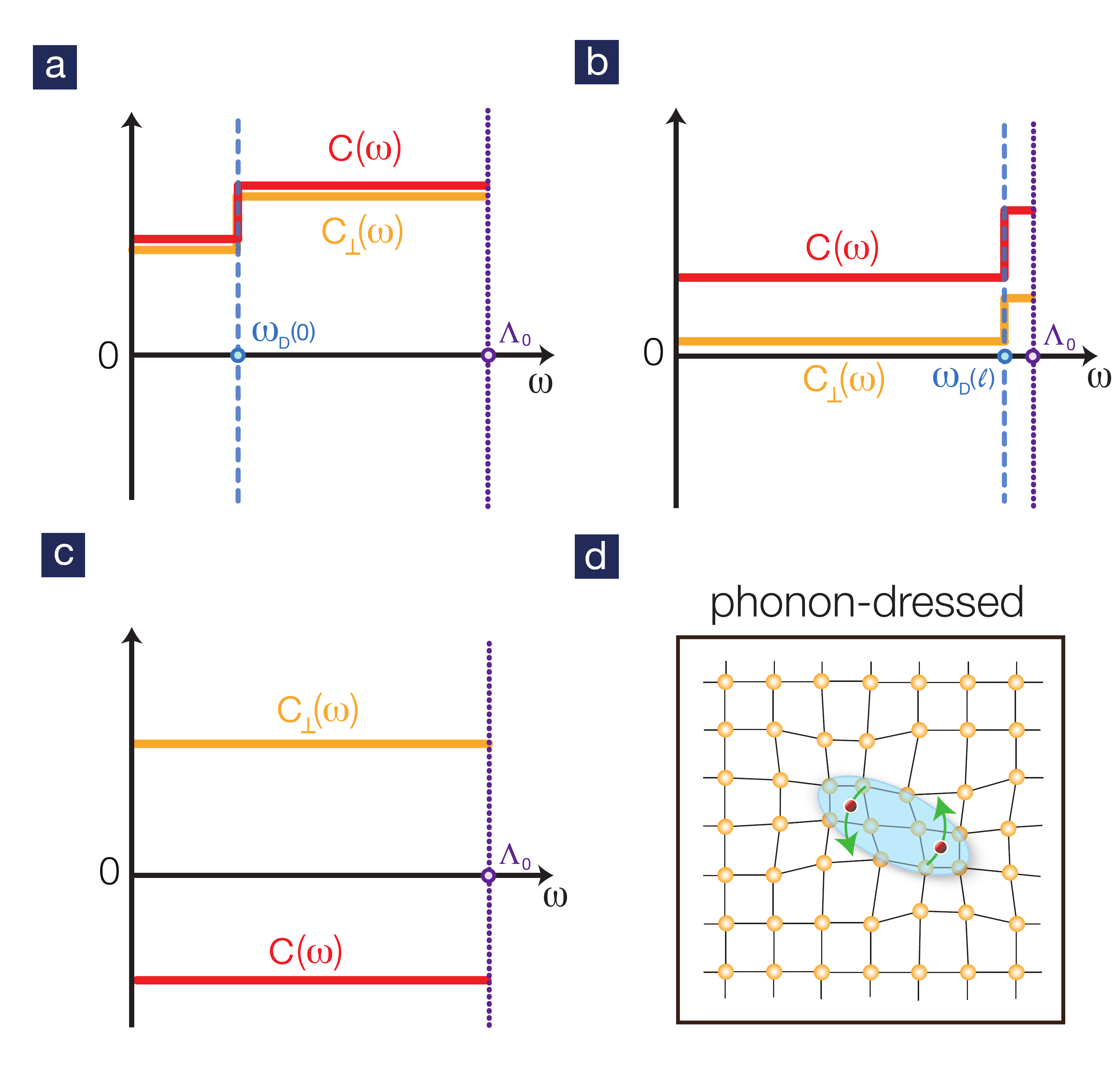}
\caption{(a) Interaction profile of the dominant intra-band $C$ and inter-band $C_\perp$ Cooper scatterings before RG transformation (b) As RG progresses, the step evolves since the Debye energy is rescaled, $\omega_D(l) = \omega_D e^l$. (c) For $l>l_D$, the distinction between simultaneous and retarded interactions disappears. (d) Schematic picture for phonon-dressed unconventional superconductor.
}
\label{dressed}
\end{figure}

By integrating the two sets of RG equations numerically, we found all couplings are well described the scaling ansatz\cite{Shih},
\begin{eqnarray}
g_i \approx \frac{G_i}{(l_d-l)^{\gamma_{g_i}}},
\qquad
\tilde{g}_i \approx \frac{\tilde{G}_i}{(l_d-l)^{\gamma_{\tilde{g}_i}}},
\end{eqnarray}
where $G_i, \tilde{G}_i$ are non-universal constants and $\gamma_{g_i}, \gamma_{\tilde{g}_i}$ are RG exponents for simultaneous and retarded couplings. The divergent length scale $l_d$, associated with the pairing gap, is solely determined by electronic origin. The dominant pairing occur within band 1 and band 2 and the Cooper scatterings $c_{11}$, $c_{22}$ $c_{12}$ have maximum exponent $\gamma_i=1$. Other Cooper scatterings are subdominant with exponents close to 0.9, as shown in Fig. 2(a). Meanwhile, by Abelian bosonization\cite{Lin97, Lin98a}, the signs of $c_{ij}$ from numerics lead to sign-revised (between electron and hole pockets) $s_{\pm}$-wave pairing, agreeing with the previous functional RG study\cite{Wang}. Note that these exponents are rather robust within the doping range where the same Fermiology maintains. What about the phonon-mediated interactions? As clearly indicated in Fig. 2(b), the RG exponents for $\tilde{c}_{11}, \tilde{c}_{22}$ are roughly 0.6, much smaller than the dominant electronic interactions, showing the pairing mechanism is electronic origin. However, since the RG exponents are positive, the retarded interactions also grow under RG transformation. These subdominant phonon-mediated interactions can lead to anomalous isotope effect as explained in the following.

To achieve quantitative understanding in weak coupling, the rescaled Debye frequency must be taken into account carefully. Under RG transformations, $\omega_D \to \omega_D e^l$ as shown in Fig. 3. At the (logarithmic) length scale $l_D \equiv \log (\Lambda_0/\omega_D)$, the difference between $g_i$ and $\tilde{g}_i$ disappears. The Debye frequency $\omega_D \sim 30$ meV in iron-based materials\cite{Boeri} and the band width (thus $\Lambda_0)$ is $3-4$ eV, giving rise to $l_D \sim 5$. Note that the RG is truncated at the cutoff length scale $l_c$ where the maximal coupling reaches order one. In weak coupling, it is clear that $l_c > l_D$ and thus the RG scheme must be divided into two steps. For $l<l_D$, both sets of RG equations are employed. At $l=l_D$, the functional form for the retarded interactions is the same as the instantaneous one. Thus, one should add up both types of couplings $g_i(l_D)+\tilde{g}_i(l_D)$ and keep running RG by just the first set of equations. In physics terms, this means that the difference between simultaneous and retarded interactions vanishes before the pairing gaps open.

\begin{figure}
\centering
\includegraphics[width=7.9cm]{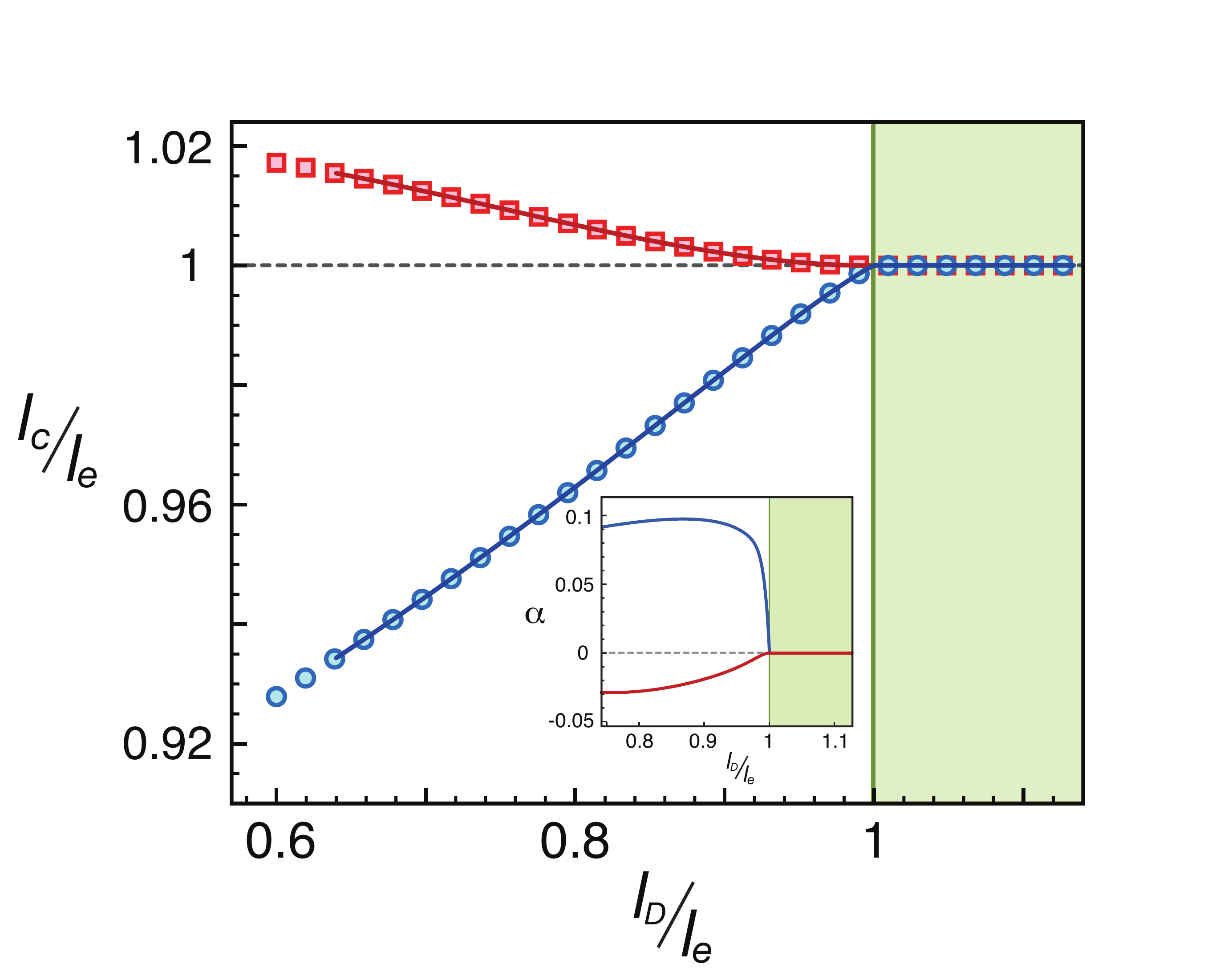}
\caption{The cutoff length scale $l_c$ versus $l_D$ for inclusion of intraband interactions $\tilde{c}_{ii}(0) = -0.3 U$ (blue circles) and interband ones $\tilde{c}_{ij}(0) = -0.14 U$ (red square), where $U$ is the strength of electron-electron interactions. For convenience, the axes are rescaled in the unit of $l_e$, the cutoff length scale with electronic interactions only. The inset shows the isotope exponent by taking numerical derivative.  
}
\label{isotope}
\end{figure}

Numerical results for the two-step RG indicate the same superconducting phase as described in previous paragraphs but the isotope exponent $\alpha$ can be extracted numerically. Since $\omega_D \sim M^{-1/2}$, it means that $d(\log M) =-2 d(\log \omega_D) = 2dl_D$. By standard scaling argument, the critical temperature takes the form, $k_BT_c\sim\Delta\left[g(0)\right] =\Delta_c e^{-l_c}$, where $\Delta_c$ is the pairing gap at the cutoff length scale. With straightforward algebra, the isotope exponent can be written as
\begin{eqnarray}\label{alpha}
\alpha=-\frac{d(\log T_c)}{d(\log M)}
\approx \frac{1}{2}\frac{d(\log\Delta_c)}{d(\log\omega_D)}+\frac{1}{2}\frac{dl_c}{dl_D}.
\end{eqnarray}
For conventional superconductor, $\Delta_c \sim \omega_D$ and the cutoff length scale is not sensitive to the Debye frequency (the second term vanishes). Thus, $\alpha \approx 1/2$. On the other hand, for unconventional superconductors without relevant electron-phonon interactions, $\Delta_c \sim \Lambda_0$ and the cutoff length scale is also not sensitive to the Debye frequency. It is clear that $\alpha=0$ in this case. 

But, what happens if the electron-phonon interactions, though not dominant, are actually relevant under RG transformation? We study how the cutoff length scale $l_c$ varies with different Debye frequencies due to isotope substitutions. In weak coupling, we found that $g_i$ are much larger than $\tilde{g}_i$. Thus, $\Delta_c$ has very weak dependence on $\omega_D$ and the first term can be ignored. The contribution from the second term is shown in Fig. 4. We tried two different profiles for the retarded interactions. Include only intra-band interactions, $\tilde{c}_{ii}(0) = -0.3 U$ first, where $U$ is the strength of electron-electron interactions. The isotope exponent is positive (reading from the slope), $\alpha \approx 0.1$, with very smooth variation. On the other hand, with only inter-band interactions, $\tilde{c}_{ii}(0) = -0.14 U$, the isotope exponent is negative and changes gradually from zero to $\alpha \approx -0.03$. These anomalous isotope effect is closely related to the pairing symmetry. For the $s_{\pm}$-wave pairing, $c_{ii}<0$ but $c_{ij}>0$ at the cutoff length scale. The phonon-mediated intra-band interactions $\tilde{c}_{ii}<0$ help to develop the pairing instability and thus lead to a positive isotope exponent. On the other hand, the inter-band ones $\tilde{c}_{ij}<0$ have opposite sign with their simultaneous counterparts $c_{ij}$. In consequence, the pairing instability is suppressed and an inverses isotope effect is in order. The RG analysis presented here provide clear and natural connection between the anomalous isotope effect and the unconventional pairing symmetry.

Although the isotope exponent $\alpha$ can be extracted numerically in weak coupling, extending the quantitative description to intermediate coupling may not be easy. If the pairing gaps open before hitting the Debye energy scale, i.e. $l_c<l_D$, our numerical results show that $l_c$ solely depend on electronic interactions and thus $dl_c/dl_D=0$. The isotope exponent in this regime mainly arises from the first term. The pairing gap $\Delta_c = \Delta_c(\Lambda_0,\omega_D e^l)$, depending on both the bandwidth and the rescaled Debye frequency, is now quite complicated. The RG analysis alone is not sufficient to obtain $\alpha$ in a quantitative fashion. However, we recently found that the effective Hamiltonian at the cutoff length scale is well captured by mean-field theory\cite{Huang2011}. In principle, one can combine RG and mean-field approaches together to compute the isotope exponent in intermediate coupling.

In summary, we investigate the competition between electronic and phonon-mediated interactions in iron-base superconductor by RG analysis. The pairing mechanism (thus pairing symmetry as well) is determined solely by electronic interactions. However, we show that the phonon-mediated retarded interactions are also relevant and give rise to anomalous isotope effect. Due to the unconventional $s_{\pm}$ pairing symmetry, the isotope exponent sensitively depends on the profile of the retarded interactions.

We acknowledge supports from the National Science Council in Taiwan through grant NSC 100-2112-M-007 -017 -MY3. Financial supports and friendly environment provided by the National Center for Theoretical Sciences in Taiwan are also greatly appreciated.

\end{document}